\documentclass[11pt,twoside]{article}

\usepackage{asp2006}
\usepackage{epsf}
\usepackage{psfig}
\usepackage{lscape}

\begin{document}
\title{Compactness of Weak Radio Sources at High Frequencies}   

\author{W. A. Majid\altaffilmark{1},E. B. Fomalont\altaffilmark{2}, D. S. Bagri\altaffilmark{1}}
\altaffiltext{1}{Jet Propulsion Laboratory, California Institute of Technology, 4800 Oak Grove Drive, Pasadena, CA 91109, U.S.A}
\altaffiltext{2}{National Radio Astronomy Observatory, 520 Edgemont Road, Charlottesville, VA 22903, U.S.A}

\begin{abstract} 

We have obtained 8.4 GHz VLBA observations of a 31-GHz complete sample
of $\sim100$ sources between 10 and 100 mJy.  The main goals of these
observations are: to determine the angular size, radio spectra and
identification for a weak sample of high frequency sources; to find
the fraction of sources which have sufficiently compact emission for
use as calibrators for VLBI observations; and for design
considerations of the proposed DSN Array.  We find that a large
fraction of observed sources have VLBI detections.  A majority of
these sources have most of their emission in a compact $<1$ mas radio
core, with remaining sources having steep radio spectra.  The source
list was provided from GBT observations to remove discrete sources in
the CBI fields.

\end{abstract}



\section{Nature of High Frequency Radio Sample}

Carrying out a VLBI survey of a complete and unbiased sample of weak
radio sources at high frequencies provide fundamental astronomical
information on the statistical and morphological properties for this
class of astrophysical objects.  Previous surveys of the nature and
structure of weak radio sources have been carried out at relatively
lower frequencies, often at 1.4 GHz \citep[e.g.][]{garrett2005}.  At
the mJy level, the proportion of AGN's are decreasing and the
population begins to be dominated by galaxies that have significant
star forming regions.  These are typically less than 3'', with about
30\% showing milliarcsecond emission \citep{muxlow2005}.

However, the angular characteristics of sources above 8 GHz are not
well-known at the mJy level.  We plan to determine the percentage of
compact milliarcsecond emission, its orientation and accurate core
position for better optical identification.  We also plan to study
spectral index correlation versus galaxy type and compare our results
with similar studies carried out for brighter sources and similar
surveys at lower frequencies.

\section{VLBI Calibrators at the mJy Scale}

Differential VLBI is routinely used to determine spacecraft positions
with accuracies of $\sim 1$ mas using compact radio sources with flux
$>$ 300 mJy at 8.4 GHz within about 5-10 degrees of the spacecraft.
Further improvement in accuracy is possible by observing calibrators
with smaller angular separation with spacecraft, reducing tropospheric
and astrometric errors.  Since sources with flux $\sim10$ mJy can be
detected with a pair of 70m antennas of the Deep Space Network (DSN),
much fainter calibrators can be used.  One of the goals of this survey
will be to determine what fraction of faint sources are useful as VLBI
calibrators.

Furthermore, NASA's DSN is considering a new generation of large
number of small ($\sim12-m$) antennas.  
The increased sensitivity of
the array along with relatively large primary beam 
may allow for in-beam phase referencing at 8 GHz if the percentage of
mJy sources with milli-arcsec components is sufficiently high.
We estimate that 30\% of such sources are compact, that is 1/5 of the
sky will contain a sufficiently strong VLBI calibrator in the primary
beam \citep{majid2007}.
At the proposed 32 GHz spacecraft telemetry, in-beam calibration will
be very rare, but a calibrator within 2 deg will permit position
determination at the 0.1 mas level.

\section{Source Selection and Summary of Observations and Data Reduction}

We compiled a list of $\sim100$ sources complete to 10 mJy at 31 GHz
from recent observations by the GBT (thanks to B.Mason, NRAO) in two
right ascension fields near the equator at 02h and 20h.  All sources
from the 1.4 GHz NRAO VLA Sky Survey (NVSS) above 6 mJy were observed
and detected by the GBT \citep{condon1998}.  We carried out 8 GHz VLBA
observations of each source with a GBT flux density $>$ 10 mJy in
January 2007 during two 10-hour observing sessions.  In a typical
phase referencing nodding style, each source was observed alternately
with a near-by phase calibrator (J2036-0629 and J0239-034).  On-source
integration time was 4 minutes, while calibrator observations were
carried out with an integration time of 30-sec.  Each target was
observed 3 times during an observing session in order to improve (u,v)
coverage.  In addition, we also included a few 30-min segments of
observations of strong ICRF sources around the sky in order to improve
the tropospheric delay model.

The NRAO Astronomical Image Processing Systems (AIPS) was used for the
data calibration and subsequent imaging.  Fringe fitting was carried
out for each phase reference calibrator.  The phase, delay and delay
rate solutions obtained were interpolated and then applied to the
program source visibility data.  The typical imaging process involving
several iterations of CLEANing and phase calibration resulted in
images with an rms thermal noise error of $\sim 1$ mJy.
Because the apriori position error of the sources were as 
large as 2" with the NVSS position, detection could only be made at
the $5-\sigma$ level over the large-field of view using the shorter VLBA
baselines.  
With an rms noise level of 1 mJy, detections could be made
at the 5 mJy level.

\section{Preliminary Results}

Of the 65 sources observed we detected 33 sources.  The majority of
detected sources are unresolved at the mas scale radio structure.
Using the JMFIT program in AIPS, we fitted each image with a
two-dimensional gaussian and determined the peak to total flux density
distribution, which we refer to as the compactness factor.  For each
source in the sample, we also obtained a spectral index using the NVSS
and GBT flux density measurements at 1.4 and 31 GHz respectively.

Figure 1 shows the distribution of the spectral index for the detected
and un-detected sample.  As expected, there is a clear trend towards
flatter spectra in the detected sample, while the undetected sample
tends to have steeper spectrum.  We also see a clear correlation in
compactness factor with the spectral index.

We attempted to identify optical counterparts for each source in
the sample.  We used the I-band digitized Palomar Schmidt archives to
search for possible counterparts down to a magnitude of 20.  Our
preliminary analysis shows that a large fraction (2/3) of compact
sources have positional association with an optical counterpart, while
only $\sim$15 \% of undetected sample indicate any association with an
optical source.

\begin{figure}[!ht]
\plotone{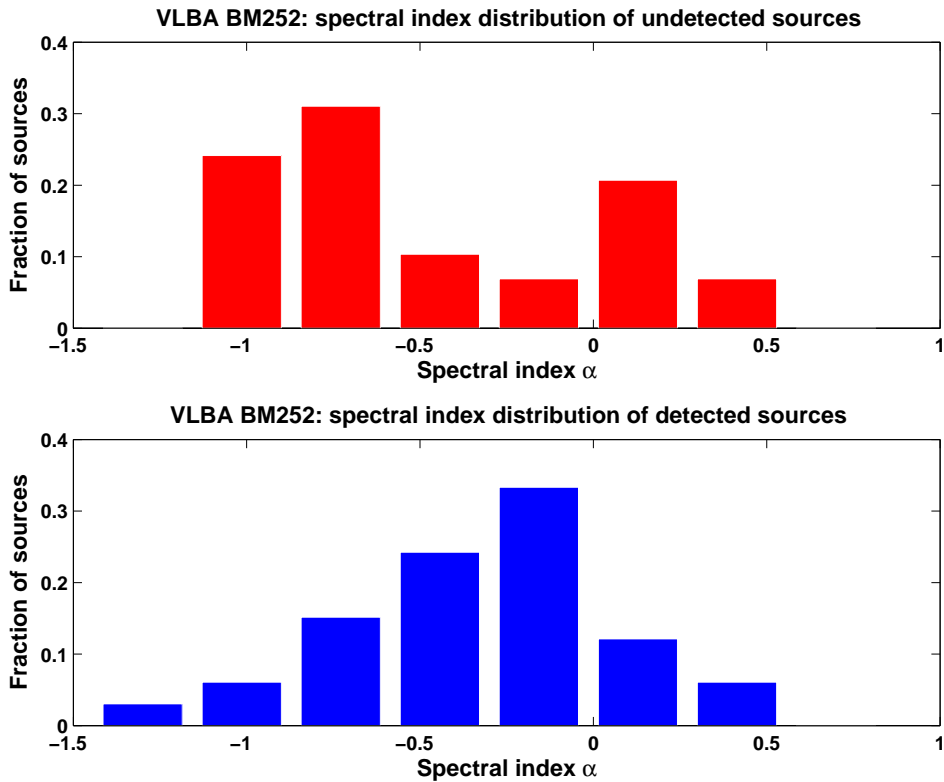}
\caption{Spectral index distribution for the undetected and
  detected samples.  The undetected sample tend to show steeper
  spectra, while the distribution of the detected sample increases
  gradually as the spectra get flatter and reaches a peak at spectral
  index of -0.2.  Spectral index is obtained from measured flux
  density at 1.4 and 31 GHz.}
\end{figure}

\section{Conclusions}

We carried out a small 8 GHz VLBA survey of a complete sample of radio
sources down to a flux density of 10 mJy.  Sources were identified
from a sample of NVSS sources with 31 GHz GBT detections.  We detected
$\sim$ 50\% of the observed sources with VLBI components.  As expected
sources with flatter spectra tend to exhibit mas emission.  In
addition, we note that the compactness factor increases for sources
with flatter spectra.  Our preliminary results also indicate a
relationship between optical identification and source compactness.

\acknowledgements 

This work was carried out at
the Jet Propulsion Laboratory, California
Institute of Technology, under contract with the National Aeronautics
and Space Administration and the National Radio Astronomy Observatory
of the National Science Foundation operated under cooperative
agreement by Associated Universities, Inc.


\end{document}